\documentclass[final,3p,times,authoryear]{elsarticle}

\makeatletter
\def\ps@pprintTitle{%
 \let\@oddhead\@empty
 \let\@evenhead\@empty
 \def\@oddfoot{\footnotesize\itshape
       \ifx\@journal\@empty Elsevier
       \else\@journal\fi\hfill\today}%
 \let\@evenfoot\@oddfoot}
\makeatother

\pdfoutput=1

\usepackage{amssymb}

\usepackage[displaymath, mathlines]{lineno}

\journal{Polar Science 9 (2), 196--207, 2015 (authors' version).
DOI: 10.1016/j.polar.2015.04.001.}

\begin{document}

\begin{frontmatter}



\title{Comparison and verification of enthalpy schemes for 
polythermal glaciers and ice sheets with a one-dimensional model}


\author[affil_ilts,affil_iaceth]{Heinz Blatter}
\author[affil_ilts]{Ralf Greve}

\address[affil_ilts]{Institute of Low Temperature Science, Hokkaido University,
Kita-19, Nishi-8, Kita-ku, Sapporo 060-0819, Japan}
\address[affil_iaceth]{Institute for Atmospheric and Climate Science,
ETH Zurich, Universit\"atstrasse 16, CH-8092 Zurich, Switzerland \\[1ex]
Corresponding author: Ralf Greve (greve@lowtem.hokudai.ac.jp)}

\begin{abstract}

The enthalpy method for the thermodynamics of polythermal glaciers and
ice sheets is tested and verified by a one-dimensional problem
(parallel-sided slab). The enthalpy method alone does not include
explicitly the transition conditions at the cold-temperate transition
surface (CTS) that separates the upper cold from the lower temperate
layer. However, these conditions are important for correctly determining
the position of the CTS. For the numerical solution of the polythermal
slab problem, we consider a two-layer front-tracking scheme as well as
three different one-layer schemes (conventional one-layer scheme,
one-layer melting CTS scheme, one-layer freezing CTS scheme). Computed
steady-state temperature and water-content profiles are verified with
exact solutions, and transient solutions computed by the one-layer
schemes are compared with those of the two-layer scheme, considered to
be a reliable reference. While the conventional one-layer scheme (that
does not include the transition conditions at the CTS) can produce
correct solutions for melting conditions at the CTS, it is more reliable
to enforce the transition conditions explicitly. For freezing
conditions, it is imperative to enforce them because the conventional
one-layer scheme cannot handle the associated discontinuities. The
suggested numerical schemes are suitable for implementation in
three-dimensional glacier and ice-sheet models.

\end{abstract}


\end{frontmatter}

\nolinenumbers


\section{Introduction}

\begin{figure}[htb]
\begin{center}
\includegraphics[width=65mm]{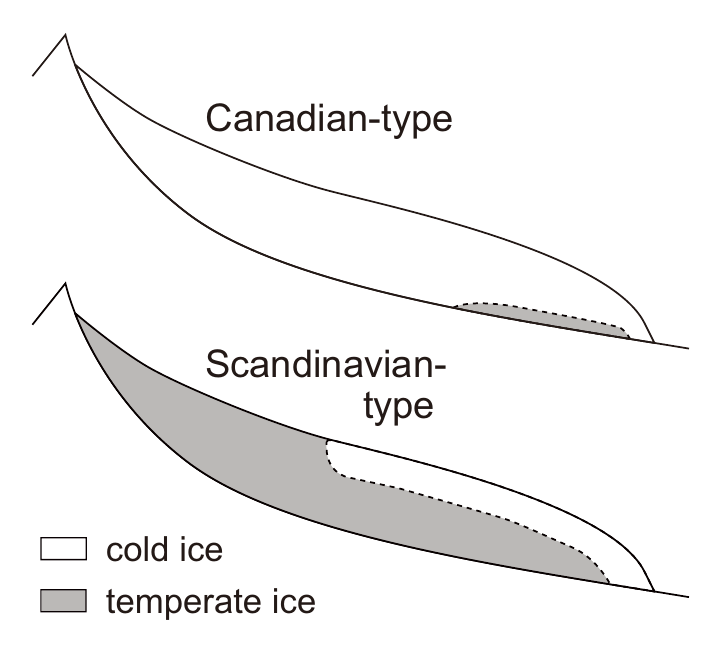}
\end{center}
\caption{Schematic cross sections of Canadian- and Scandinavian-type
polythermal glaciers \citep[adapted from][]{AschwandenBueler2012}.}
\label{fig:canadian_scandinavian}
\end{figure}

The decrease of the ice viscosity with increasing content of liquid
water in temperate ice was first confirmed and measured by
\citet{Duval1977}. It is therefore desirable to simulate the water
content in glaciers and ice sheets realistically, especially if the
temperate ice occurs in a basal layer where shear deformation is
largest. Mathematical models of polythermal ice masses were introduced
and further developed by \citet{Fowler1978}, \citet{Hutter1982},
\citet{Fowler1984} and \citet{Hutter1993}. We distinguish essentially
two types of polythermal glaciers, Canadian-type polythermal glaciers,
which are cold in most of the ice mass except for a temperate basal
layer in the ablation zone, and Scandinavian-type glaciers, which are
temperate in most parts except for a cold surface layer in the ablation
zone (Fig.~\ref{fig:canadian_scandinavian}) \citep{Blatter1991}.

This work attempts to verify thermodynamic schemes used in shallow ice
sheet models. Therefore, we do not investigate processes which are not
usually included in ice sheet models, such as possible diffusion of
water in temperate ice \citep{Hutter1993} and pre-melting in ice at
sub-freezing temperatures \citep{Dash&al2006}. For verification of
numerical solutions with exact solutions, we neglect the pressure
dependence of the melting point and the temperature dependence of the
heat conductivity and specific heat capacity. Following
\citet{Aschwanden2005}, ``ice is treated as temperate if a change in
heat content leads to a change in liquid water content alone, and is
considered cold if a change in heat content leads to a temperature
change alone.'' This implies that temperate ice is at the melting point
and the temperatures in cold ice are below the melting point.

Polythermal schemes that solve the field equations for the cold and
temperate layers separately were implemented for both types of
polythermal glaciers, in one dimension for the Scandinavian-type
Storglaci\"aren, Sweden \citep{Pettersson2007}, in two dimensions for
the Canadian-type Laika Glacier, Canada \citep{Blatter1991} and for
three-dimensional ice sheets, which are Canadian-type polythermal
\citep{Greve1997}. With the assumption that water mostly accumulates
along the trajectories of ice particles in the temperate layer,
\citet{Aschwanden2005} used a trajectory model to determine the position
of the cold-temperate transition surface (CTS) and the water content
in the temperate part of Storglaci\"aren. \citet{AschwandenBueler2012}
suggested an enthalpy scheme with the idea that, with enthalpy, only one
thermodynamic field variable must be computed, and the temperature and
water content result from the enthalpy as diagnostic fields. The
domains of cold and temperate ice are discriminated by the contour of
the enthalpy of ice with no liquid water content at the melting point.

A crucial point in polythermal enthalpy schemes is their treatment of
the Stefan-type energy- and mass-flux matching conditions at the CTS,
which are important for determining its position \citep{Greve1997}.
These transition conditions are not included explicitly in the
formulation of the enthalpy scheme according to
\citet{AschwandenBueler2012}.

Two different cases must be distinguished. Melting conditions occur if
cold ice flows across the CTS into the temperate region. At the CTS, the
particles consist of ice at melting temperature without liquid water,
and, after the transition, start to accumulate water due to strain
heating. Thus, the boundary condition on the temperate side of the CTS
is zero water content. To match the vanishing latent heat flux, the
diffusive heat flux and corresponding enthalpy gradient on the cold side
must also vanish.

The situation is different for freezing conditions at the CTS, where the
ice flows from the temperate region into the cold region and the liquid
water content of the temperate ice freezes at the CTS. The advective
latent heat flux on the temperate side then changes into a diffusive
heat flux on the cold side. Thus, a drop of a non-vanishing water
content to zero results in a non-vanishing temperature (enthalpy)
gradient in the cold layer at the CTS.

This work attempts to verify and test modified enthalpy methods, and in
particular to test how the modified schemes handle the internal boundary
between cold and temperate ice. For the verification, we use an exact
solution which is available for steady states in a parallel-sided slab,
which reduces the problem to one dimension \citep{Greve1997,
GreveBlatter2009}. In Section~\ref{sec:enthalpy}, we review the main
concepts of the enthalpy method, and in Section~\ref{sec:slab}, we
formulate the enthalpy method for the special case of the parallel-sided
slab. Section~\ref{sec:numerics} deals with different one- and two-layer
methods to solve this problem, the two-layer front-tracking scheme being
used to provide reference solutions against which the performance of the
simpler one-layer methods can be tested. Concrete numerical experiments
are defined in Section~\ref{sec:setup}, and results are presented and
discussed in Sections~\ref{sec:results} and \ref{sec:discussion}.

\section{Enthalpy formulation}
\label{sec:enthalpy}

In this paper we follow the formulation of \citet{AschwandenBueler2012}
and use the notation of \citet{GreveBlatter2009}. All physical
parameters, namely the stress exponent $n=3$ and the rate factor
$A=5.3\times{}10^{-24}\,\mathrm{s^{-1}\,Pa^{-3}}$ of Glen's flow law,
the heat conductivity of ice,
$\kappa=2.1\,\mathrm{W\,m^{-1}\,K^{-1}}$, the melting point of ice,
$T_\mathrm{m}=0^\circ\mathrm{C}$, the density of ice,
$\rho=910\,\mathrm{kg\,m^{-3}}$, the specific heat capacity of ice,
$c=2009\,\mathrm{J\,kg^{-1}\,K^{-1}}$, and the latent heat of fusion,
$L=3.35\times{}10^5\,\mathrm{J\,kg^{-1}}$, are assumed to be constant
for simplicity. In particular, this means that we neglect the slightly
larger density of temperate ice compared to cold ice due to the
$\sim\!10\%$ larger density of liquid water.

The liquid water content (mass fraction) of temperate ice is defined by
\begin{equation}
  W = \frac{\rho_\mathrm{w}}{\rho}\,,
\end{equation}
where $\rho_\mathrm{w}$ is the partial density of liquid water in the
mixture. Let $h_\mathrm{m}=c\,T_\mathrm{m}$ be the enthalpy of ice at
the melting temperature with vanishing water content. For cold ice with
a temperature $T$ and temperate ice with a water content $W$, the
enthalpy is then given by
\begin{equation}
  \label{eq:enthalpy-definition-cases}
  h = \left\{ \begin{array}{ll}
         c T\,,                      & T < T_\mathrm{m}\,, \\
         h_\mathrm{m} + L W\,, \quad & T = T_\mathrm{m} 
                                       \;\mbox{ and }\; 0 \le W < 1\,,
     \end{array} \right.
\end{equation}
and the corresponding balance equation reads
\begin{equation}
  \label{eq:enthalpy-full}
  \rho \left(
  \frac{\partial h}{\partial t} + \mathbf{v}\cdot\mathrm{grad}\,h \right)
  = -\nabla \cdot \mathbf{q} + \mathrm{tr}(\mathbf{t}\cdot\mathbf{D})\,,
\end{equation}
where $t$ is time, $\mathbf{v}$ the velocity vector,
$\mathbf{q}$ the heat flux vector, $\mathbf{t}$ the 
Cauchy stress tensor, $\mathbf{D}$ the strain-rate tensor,
the middle dot $(\cdot)$ denotes tensor contraction
and tr denotes the trace of a tensor. The heat flux is given by
the constitutive equation 
\begin{equation}
  \label{eq:diffusive_heat_flux}
  \mathbf{q} = - \frac{\kappa_\mathrm{c,t}}{c}\,\nabla h\,,
\end{equation}
with the conductivity
\begin{equation}
  \label{eq:conductivity}
  \kappa_\mathrm{c,t} = 
     \left\{ \begin{array}{ll}
         \kappa\,, \quad & h < h_\mathrm{m}\,, \\
         0\,,            & h \geq h_\mathrm{m}\,.
     \end{array} \right.
\end{equation}
For cold ice ($h<h_\mathrm{m}$), this is Fourier's law of heat
conduction, while for temperate ice ($h\ge{}h_\mathrm{m}$), the heat
flux is omitted because of the vanishing temperature gradient ($\nabla
T_\mathrm{m}=0$) and the negligibly small (at least for small water
content, $W \ll 1$) water diffusion.

\section{Polythermal slab}
\label{sec:slab}

\begin{figure}[htb]
\begin{center}
\includegraphics[width=80mm]{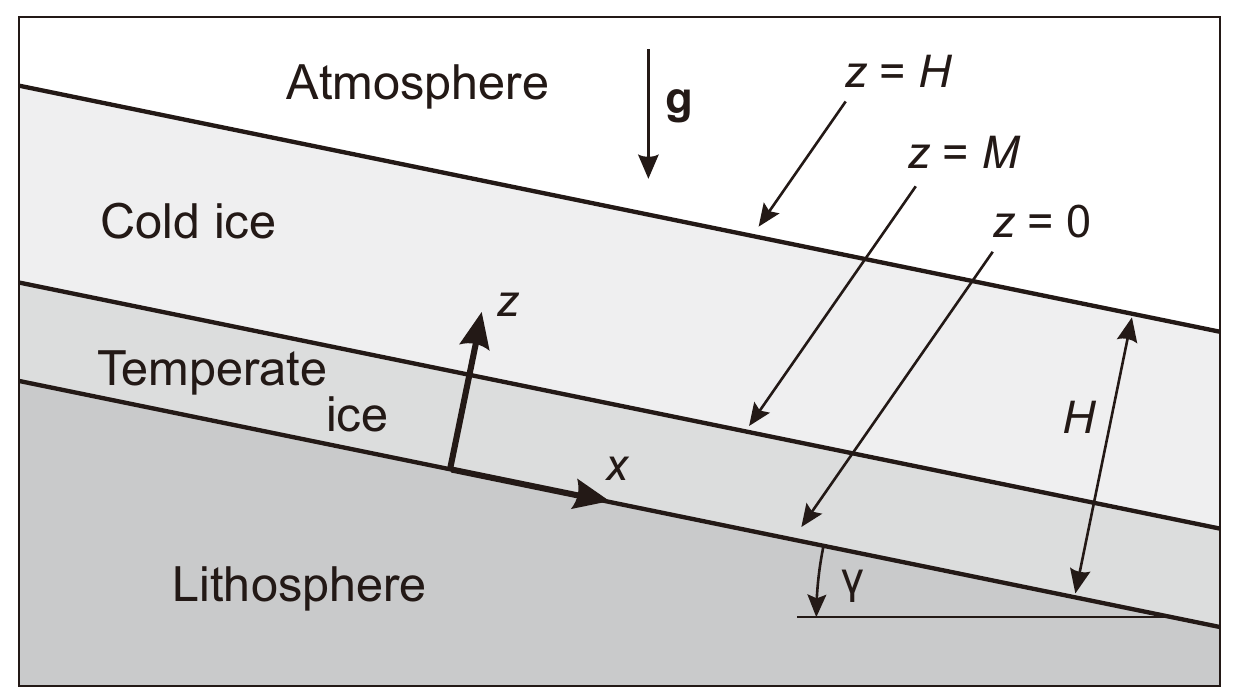}
\end{center}
\caption{Polythermal parallel-sided ice slab geometry and coordinate
system \citep[adapted from][]{GreveBlatter2009}.}
\label{fig:twolayerslab}
\end{figure}

To reduce the problem of a polythermal ice mass to a one-dimensional
problem, we apply the plane strain approximation for a two-dimensional
flow in the vertical $x$-$z$ plane of a parallel-sided slab with
constant and steady thickness $H$ and constant inclination angle
$\gamma$, and without dependencies on the transverse $y$ coordinates
(Fig.~\ref{fig:twolayerslab}). Furthermore, we impose uniformity in the
down-slope ($x$) direction, that is, $\partial\psi/\partial{}x=0$ for
any field quantity $\psi$. Thermomechanical coupling is
omitted, thus strain heating due to horizontal shearing is prescribed
\citep[e.g.,][]{Greve1997, GreveBlatter2009},
\begin{equation}
  \label{eq:strainheating}
  \mathrm{tr}(\mathbf{t}\cdot\mathbf{D}) 
  = 2A\,(\rho g \sin\gamma)^4\,(H-z)^4\,.
\end{equation}

Otherwise, the downslope velocity profile $v_x(z)$ and basal sliding
are not relevant for the problem in consideration. Owing to the assumed
uniformity in $x$-direction and the plane strain approximation, the
continuity equation (mass balance) takes the form
$\partial{}v_z/\partial{}z=0$ for the slab problem, so that the
velocity component in $z$-direction, $v_z$, is constant over depth,
$v_z = \mathrm{const}$. Due to the kinematic boundary conditions,
this means that the accumulation/ablation rate at the slab surface
is equal to the melting/freezing rate at the base (and both are
equal to $-v_z$).

With all these conditions, and neglect of water diffusion, the balance
equation for enthalpy, Eq.~(\ref{eq:enthalpy-full}), is reduced to
\begin{equation}
  \label{eq:field-enthalpy-time}
  \frac{\partial h}{\partial t}
  = -v_z \frac{\partial h}{\partial z} + 
  \frac{1}{\rho\,c}\, \frac{\partial}{\partial z}
  \left(\kappa_\mathrm{c,t}\, \frac{\partial h}{\partial z} \right) + 
  \frac{2A}{\rho}\,(\rho g \sin\gamma)^4(H-z)^4\,.
\end{equation}
We restrict this study to the Canadian-type polythermal slab. The
imposed enthalpy boundary condition at the surface of the cold layer is
\begin{equation}
  h|_{z=H} = h_\mathrm{s}(t) = c\,T_\mathrm{s}(t)\,.
  \label{eq:4.12}
\end{equation}
Depending on the direction of the ice flow through the CTS, melting and
freezing conditions must be distinguished. Let
\begin{equation}
  \label{eq:volumeflux}
  a_\mathrm{m}^\perp = w_z - v_z
\end{equation} 
be the ice volume flux relative to the moving CTS, where
$w_z=\mathrm{d}M/\mathrm{d}t$ is the kinematic (migration) velocity of
the CTS in $z$-direction. Melting conditions are then characterised by
$a_\mathrm{m}^\perp > 0$ and freezing conditions by $a_\mathrm{m}^\perp
< 0$. We limit our considerations to the case that the kinematic
velocity of the CTS is smaller than the ice velocity in $z$-direction
($|w_z|<|v_z|$), so that melting and freezing conditions correspond to
downward ($v_z<0$) and upward ice motion ($v_z>0$), respectively.

For melting conditions, ice at the melting temperature with
no liquid water flows from the cold into the temperate layer,
where water is produced along a trajectory of an ice particle by strain
heating, thus
\begin{equation}
  \label{eq:meltcond}
  h^+ = h^- = h_\mathrm{m}\,\quad (W^-=0)\,,
  \qquad
  {\left(\frac{\partial h}{\partial z}\right)\!\!}^+ = 0
\end{equation}
(where the cold layer is defined as the positive, and the temperate layer
as the negative side of the CTS).
Together with the imposed surface enthalpy $h_\mathrm{s}$, the enthalpy
$h_\mathrm{m}$ at the CTS and the vanishing enthalpy gradient constitute
three boundary conditions for the cold layer, which determine the
evolution of the enthalpy profile and the thickness of the cold layer,
thus also the position of the CTS. Since we have neglected water
diffusion in the temperate layer, the enthalpy $h_\mathrm{m}$ at the CTS
alone determines the evolution of the enthalpy profile in the temperate
layer; no additional basal boundary condition is required. However,
if a regularising small water diffusion were applied, an additional
basal boundary condition would be needed, which then should be chosen
carefully in order not to influence the numerical solution significantly
(for instance, a zero-flux condition).

For freezing conditions, the enthalpy released by freezing of water
flows into the cold ice along the enthalpy gradient,
\begin{equation}
  h^+ = h_\mathrm{m}\,,
  \qquad
  \frac{\kappa}{\rho c} 
    {\left(\frac{\partial h}{\partial z}\right)\!\!}^+
  = (h^--h_\mathrm{m})\, a_\mathrm{m}^\perp\,.
  \label{eq:freezcond}
\end{equation}
With $a_\mathrm{m}^\perp < 0$, $(\partial{}h/\partial{}z)^+$ can be
strictly negative and $(h^--h_\mathrm{m})$ strictly positive (equivalent
to $W^->0$, discontinuous water content at the CTS). The advection
equation for enthalpy in the temperate layer requires one boundary
condition, for which we use the enthalpy at the ice base. Assuming
a vanishing water content of the basal ice yields
\begin{equation}
  h|_{z=0} = h_\mathrm{m}\,.
  \label{eq:enth_ice_base}
\end{equation}
For the cold layer, the Dirichlet conditions (\ref{eq:4.12})
(prescribed surface enthalpy) and (\ref{eq:freezcond})$_1$
determine the evolution of the enthalpy profile, and 
condition (\ref{eq:freezcond})$_2$ can be used to determine
the evolution of the CTS position by solving it for the kinematic
velocity $w_z$ that is contained in the volume flux
$a_\mathrm{m}^\perp$.

Equation (\ref{eq:field-enthalpy-time}) with $\kappa_\mathrm{t}=0$ for
the temperate layer is equivalent to the accumulation of water produced
by strain heating along trajectories. As stated above, we only consider
the case $|w_z|<|v_z|$, so that melting (freezing) conditions correspond
to downward (upward) ice motion. Therefore, the water content tends to
increase downwards from the CTS for melting conditions and upwards from
the bed for freezing conditions.

Equations (\ref{eq:field-enthalpy-time})--(\ref{eq:enth_ice_base}), to
be complemented by an initial enthalpy profile $h_\mathrm{init}$,
constitute the problem of the Canadian-type polythermal parallel-sided
slab. Under the additional assumption of steady-state conditions
($\partial\psi/\partial{}t=0$ for any field quantity $\psi$), they can
be solved exactly, with the exception of the position of the CTS that
must be obtained with a numerical root-finding algorithm (for details
see \citet{Greve1997} or \citet{GreveBlatter2009}). For the general,
transient case, a numerical solution is required.

\section{Numerical methods}
\label{sec:numerics}

An established strategy to compute polythermal ice masses numerically is
to split the computational domain into two distinct domains of cold and
temperate ice, and compute the respective temperature and water content
on two different grids \citep{Blatter1991, Greve1997, Pettersson2007}.
However, for the sake of simplicity, it is desirable to produce a
numerical solution of the problem with a one-layer scheme, i.e., on one
grid that spans the entire polythermal domain. Here, we describe
possibilities for enthalpy-based two-layer and one-layer schemes for
both melting and freezing conditions at the CTS.

For all schemes, time is discretized by
\begin{equation}
  t_n = t_\mathrm{init} + n \Delta{}t
  \quad
  (n = 0 \ldots n_\mathrm{max})\,, 
\end{equation}
where $t_\mathrm{init}$ is the initial time of the respective
simulation, $n$ is the discrete time index and $\Delta{}t$ the time
step. We have developed explicit (Euler-forward) and implicit
(Euler-backward) versions of all schemes, based on second-order centred
finite differences for the first and second derivatives with respect to
$z$ in the diffusion-advection equation in the cold layer
(Eq.~(\ref{eq:field-enthalpy-time}) with $\kappa_\mathrm{c}=\kappa$),
and upstream first-order differences for the first derivatives in the
advection equation in the temperate layer
(Eq.~(\ref{eq:field-enthalpy-time}) with $\kappa_\mathrm{t}=0$).

\subsection{Two-layer front-tracking scheme}
\label{ssec:numerics_two_layer}

The surface of the cold layer is at $z=H$, the bottom of the cold and
top of the temperate layer (thus the CTS) at $z=M(t)$, and the bottom of
the temperate layer at $z=0$ (Fig.~\ref{fig:twolayerslab}). To solve
the enthalpy equation (\ref{eq:field-enthalpy-time}), we map both
layers separately to layers of thickness unity,
\begin{equation}
  \zeta_\mathrm{c} = \frac{z - M(t)}{H - M(t)}\ ,\quad \zeta_\mathrm{t}
  = \frac{z}{M(t)}\ ,\quad \tau = t \,,
\end{equation}
where $\zeta_\mathrm{c}$ and $\zeta_\mathrm{t}$ are the transformed
vertical coordinates in the cold and temperate layer, respectively,
and $\tau$ is the transformed time. The CTS is therefore fixed with the
lower and upper boundaries of the cold and temperate domains
($\zeta_\mathrm{c}=0$ / $\zeta_\mathrm{t}=1$), respectively. The
transformed equations (\ref{eq:field-enthalpy-time}) are for the
cold layer
\begin{eqnarray}
  \label{eq:cold}
  \frac{\partial h}{\partial \tau} &=& 
  \frac{w_z\, (1-\zeta_\mathrm{c})-v_z}{H-M}\,
  \frac{\partial h}{\partial \zeta_\mathrm{c}} + 
  \frac{\kappa}{\rho\,c}\,\frac{1}{(H-M)^2}\,
  \frac{\partial^2 h}{\partial \zeta_\mathrm{c}^2} \nonumber\\
  && +\,\frac{2A}{\rho}(\rho g \sin\gamma)^4(H-M)^4 (1-\zeta_\mathrm{c})^4\,,
\end{eqnarray}
and for the temperate layer
\begin{eqnarray}
  \label{eq:temperate}
  \frac{\partial h}{\partial \tau} = 
  \frac{w_z\, \zeta_\mathrm{t}-v_z}{M}\, 
  \frac{\partial h}{\partial \zeta_\mathrm{t}} +
  \frac{2A}{\rho}(\rho g \sin\gamma)^4 (H-M \zeta_\mathrm{t})^4\,,
\end{eqnarray}
where $w_z=\mathrm{d}M/\mathrm{d}t$ is the kinematic velocity of
the CTS introduced in Section~\ref{sec:slab}.

The spatial grids for the cold and temperate domain are defined by
\begin{equation}
  (\zeta_\mathrm{c})_{k_\mathrm{c}}
  = \frac{k_\mathrm{c}}{k_\mathrm{c,max}}
  \quad
  (k_\mathrm{c} = 0 \ldots k_\mathrm{c,max})
\end{equation}
and
\begin{equation}
  (\zeta_\mathrm{t})_{k_\mathrm{t}}
  = \frac{k_\mathrm{t}}{k_\mathrm{t,max}}
  \quad
  (k_\mathrm{t} = 0 \ldots k_\mathrm{t,max})\,, 
\end{equation}
where $k_\mathrm{c}$ and $k_\mathrm{t}$ are the discrete grid indices
for the two domains, and $k_\mathrm{c,max}$ and $k_\mathrm{t,max}$ denote
the number of grid points in the domains. For each time step from
a given time $t_n$ to the new time $t_{n+1}=t_n+\Delta{}t$,
Eqs.~(\ref{eq:cold}) and (\ref{eq:temperate}) are solved on these grids
with the discretizations described above.

For melting conditions at the CTS, the enthalpy $h_\mathrm{s}$ at the
surface, the enthalpy $h_\mathrm{m}$ and the enthalpy gradient on the
cold side of the CTS are defined, thus the position of the CTS is
determined by the enthalpy profile in the cold layer alone. With the two
boundary conditions for the cold layer, given surface enthalpy
$h_\mathrm{s}$ and given enthalpy $h_\mathrm{m}$ at the given CTS,
$z=M_n$, obtained for $t_n$, an integration step
of Eq.~(\ref{eq:cold}) does not generally result in a vanishing enthalpy
gradient at $M_n$. By approximating the enthalpy profile around $M_n$ by
a quadratic parabola, the position of the vertex of the parabola is a
first approximation $M^{(1)}_{n+1}$ of the CTS position at the new time
$t_{n+1}$. The position can then be iterated (iteration index $i$)
to the desired accuracy,
\begin{equation}
  M^{(i+1)}_{n+1} = M^{(i)}_{n+1} -
            {\left(\frac{\partial h}{\partial z}\right)\!\!}^+
            \,\Bigg|_{n+1}^{(i)}
            \;\Big/\;
            {\left(\frac{\partial^2 h}{\partial z^2}\right)\!\!}^+
            \,\Bigg|_{n+1}^{(i)}\,,
\end{equation}
using the first and second derivative of the enthalpy profile at
$M^{(i)}_{n+1}$. Furthermore, from the displacement of the CTS,
we obtain the kinematic velocity $(w_z)_{n+1}$ of the CTS via
\begin{equation}
  (w_z)_{n+1} = \frac{M_{n+1}-M_{n}}{\Delta{}t}\,.
\end{equation}

For freezing conditions, the transition condition at the CTS
(Eq.~(\ref{eq:freezcond})$_2$) in the transformed cold layer yields an
equation for the ice volume flux $a_\mathrm{m}^\perp$ through the CTS,
\begin{equation}
  \label{eq:jump2}
  a_\mathrm{m}^\perp
  = \frac{\kappa}{(h^--h_m)\,\rho c\, (H-M)}\,
    {\left(\frac{\partial h}{\partial\zeta}\right)\!\!}^+\,.
\end{equation}
The discretized version of this equation provides the volume flux
$(a_\mathrm{m}^\perp)_{n+1}$ and, via Eq.~(\ref{eq:volumeflux}), the
kinematic velocity of the CTS,
\begin{equation}
  \label{eq:velcts}
  (w_z)_{n+1} = (a_\mathrm{m}^\perp)_{n+1} + v_z\,,
\end{equation}
which allows to update the CTS position,
\begin{equation}
  \label{eq:cts_position_update}
  M_{n+1} = M_n + (w_z)_{n+1} \Delta{}t\,.
\end{equation}

\subsection{Conventional one-layer scheme}
\label{ssec:numerics_one_layer_1}

In contrast to the two-layer scheme discussed in
Section~\ref{ssec:numerics_two_layer}, in the conventional
one-layer scheme, which corresponds to the enthalpy scheme by
\citet{AschwandenBueler2012}, Eq.~(\ref{eq:field-enthalpy-time})
is solved for the entire polythermal slab on one grid in the
$z$-domain. It is defined by
\begin{equation}
  z_k = k \Delta{}z = H \frac{k}{k_\mathrm{max}}
  \quad
  (k = 0 \ldots k_\mathrm{max})\,,
\end{equation}
where $k$ is the discrete grid index, $\Delta{}z$ the grid spacing
(resolution), and $k_\mathrm{max}$ denotes the number of grid points.
This grid is also used for the modified one-layer schemes described
below (Sections~\ref{ssec:numerics_one_layer_2},
\ref{ssec:numerics_one_layer_3}).

The CTS must be tracked on this grid. The cold and temperate layers can
be discriminated by the contour $h=h_\mathrm{m}$, and we define the CTS
position, $k=k_\mathrm{cts}$, as the uppermost grid point of the
temperate part (that is, the uppermost grid point for which 
$h\ge{}h_\mathrm{m}$ holds).

Transition conditions at the CTS are not accounted for explicitly. In
our one-dimensional implementation, for a melting CTS, where the ice
flows downwards, the surface boundary condition (given enthalpy) and the
assumed continuity of the enthalpy field at the CTS define the entire
enthalpy profile $h_{k,n+1}$ at the new time $t_{n+1}$ when the profile
$h_{k,n}$ at the old time $t_n$ is known. Boundary conditions at the
base are not necessary because of the advection equation in the
temperate layer. Therefore, only one boundary condition at the surface
of the cold layer is required to obtain a unique solution.

Although the conductivity is constant in the cold and temperate layers
($\kappa_\mathrm{c}=\kappa$ and $\kappa_\mathrm{t}=0$, respectively;
Eq.~(\ref{eq:conductivity})), the discretization of the diffusion
term in Eq.~(\ref{eq:field-enthalpy-time}) must take into account the
variation of the conductivity at least across the CTS, i.e., for
$k=k_\mathrm{cts}$:
\begin{eqnarray}
  \frac{\partial}{\partial z}
  \left(\kappa_\mathrm{c,t}\,\frac{\partial h}{\partial z}\right)
  &\sim&
  \frac{(\kappa_\mathrm{c,t}\,\frac{\partial h}{\partial z})_{k+1/2,n}
        -(\kappa_\mathrm{c,t}\,\frac{\partial h}{\partial z})_{k-1/2,n}}
       {\Delta{}z}
  \nonumber\\[2ex]
  &\sim&
  \frac{\kappa_\mathrm{c}(h_{k+1,n}-h_{k,n})
        -\kappa_\mathrm{t}(h_{k,n}-h_{k-1,n})}
       {\Delta{}z^2}
  \label{eq:discr_across_cts}
\end{eqnarray}
(T.\ Kleiner, personal communication, February 2014;
\citeauthor{Kleiner&al2015}, \citeyear{Kleiner&al2015}).
Omission of this, and discretizing the diffusion term in the form
$\kappa_\mathrm{c,t}\,(\partial^2{}h/\partial{}z^2)$ instead,
results in a faulty enthalpy profile that violates the
melting-CTS transition condition (\ref{eq:meltcond})$_2$ (zero
enthalpy gradient at the cold side of the CTS).

For freezing conditions, the method must fail because it is not
consistent with the discontinuity of the enthalpy field at the CTS that
results from the discontinuity of the water content
(Eq.~(\ref{eq:freezcond}) and following text).

\subsection{One-layer melting CTS scheme}
\label{ssec:numerics_one_layer_2}

We propose an alternative one-layer scheme for melting conditions that
enforces explicitly the zero enthalpy gradient at the cold side of the
CTS (Eq.~(\ref{eq:meltcond})$_2$). For this scheme, the discretization
of the diffusion term in Eq.~(\ref{eq:field-enthalpy-time}) need not
consider a conductivity that depends on the position $z$ like in
Eq.~(\ref{eq:discr_across_cts}); instead, it can be done in the form
\begin{equation}
  \begin{array}{l}
    \kappa_\mathrm{c,t}\,
    \mbox{$\displaystyle\frac{\partial^2 h}{\partial z^2}$}
    \,\sim\,
    (\kappa_\mathrm{c,t})_{k,n}\,
    \mbox{$\displaystyle\frac{h_{k+1,n} - 2h_{k,n} 
                              + h_{k-1,n}}{\Delta{}z^2}$}\,,
    \\[4ex]
    \hspace*{5.3em}\mbox{with }
    (\kappa_\mathrm{c,t})_{k,n} =
    \left\{ \begin{array}{ll}
      \kappa_\mathrm{t}\,, \quad & k = 0 \ldots k_\mathrm{cts}\,,
      \\
      \kappa_\mathrm{c}\,, \quad & k = k_\mathrm{cts}+1 \ldots k_\mathrm{max}\,.
    \end{array} \right.
  \end{array}
  \label{eq:discr_const_kappa}
\end{equation}

Each time step is now divided into two iteration steps. The
predictor step for solving the enthalpy equation (\ref{eq:field-enthalpy-time})
is carried out for the entire polythermal slab as in the conventional
one-layer scheme (Section~\ref{ssec:numerics_one_layer_1}). This provides
a preliminarily updated enthalpy profile $h^{(1)}_{k,n+1}$.
For this profile, the updated position of the CTS 
($k=k_\mathrm{cts}$, uppermost temperate grid point) and the updated
conductivities $\kappa_\mathrm{c,t}$ (according to
Eq.~(\ref{eq:conductivity})) for each grid point are determined.

The corrector step affects only the cold layer. We repeat the
forward step for the enthalpy equation from the grid point
$k=k_\mathrm{cts}$ to the surface ($k=k_\mathrm{max}$), discretizing the
zero enthalpy gradient on the cold side of the CTS
(Eq.~(\ref{eq:meltcond})$_2$) by
\begin{equation}
  h_{k_\mathrm{cts},n+1} = h_{k_\mathrm{cts}+1,n+1} \,.
\end{equation}
This provides an enthalpy profile $h^{(2)}_{k,n+1}$ for the cold layer
only. The complete, updated enthalpy profile $h_{k,n+1}$ is assembled by
the predictor step for the temperate layer and the corrector step for the
cold layer,
\begin{equation}
  h_{k,n+1} =
  \left\{\begin{array}{ll}
    h^{(1)}_{k,n+1}\,,
    \quad &
    k = 0 \ldots k_\mathrm{cts}\,,
    \\[2ex]
    h^{(2)}_{k,n+1}\,,
    \quad &
    k = k_\mathrm{cts}+1 \ldots k_\mathrm{max}\,.
  \end{array}\right.
  \label{eq:enthalpy_profile_assembly}
\end{equation}

\subsection{One-layer freezing CTS scheme}
\label{ssec:numerics_one_layer_3}

Owing to the discontinuous enthalpy and enthalpy gradient at the CTS
(Eq.~(\ref{eq:freezcond})), a one-layer scheme for freezing conditions
at the CTS is more difficult to implement. The advection equation in the
temperate layer (Eq.~(\ref{eq:field-enthalpy-time}) with
$\kappa_\mathrm{t}=0$) only requires one boundary condition, which is
the Dirichlet condition (\ref{eq:enth_ice_base}) (imposed value of the
basal enthalpy corresponding to zero water content). The advection
equation can therefore be solved independently of the cold layer. For
each time step (from $t_n$ to $t_{n+1}$), we do so for the entire
polythermal slab, which provides a first profile $h^{(1)}_{k,n+1}$ that
is only valid for the temperate layer from the ice base to the yet 
unknown new position of the CTS.

We then solve the diffusion-advection equation
(Eq.~(\ref{eq:field-enthalpy-time}) with $\kappa_\mathrm{c}=\kappa$) in
the cold layer. Based on the CTS position $M_n$ at time $t_n$ (in
contrast to the one-layer melting CTS scheme, the CTS is tracked with
sub-grid precision, as will be explained below), we denote the uppermost
grid point in the temperate layer as $k_\mathrm{cts}$, and compute the
enthalpy profile from the grid point $k=k_\mathrm{cts}$ to the surface
($k=k_\mathrm{max}$). A quadratic extrapolation using the three grid
points above $k_\mathrm{cts}$ is used as a boundary condition:
\begin{equation}
  h_{k_\mathrm{cts},{n+1}}
  = 3\,h_{k_\mathrm{cts}+1,{n+1}}
    - 3\,h_{k_\mathrm{cts}+2,{n+1}} 
    +  h_{k_\mathrm{cts}+3,{n+1}}\,.
\end{equation}
This step provides a second enthalpy profile $h^{(2)}_{k,n+1}$ that is
valid for the cold layer from the CTS to the surface. 

Tracking of the CTS is carried out by using the transition condition
(\ref{eq:freezcond})$_2$. For this purpose, the enthalpy on the
temperate side of the CTS is interpolated by
\begin{equation}
  h^-_{n+1} 
  \sim h^{(1)}_{k_\mathrm{cts}}
       + \frac{M_n-(z_{k_\mathrm{cts}})_n}{\Delta z}\,
         \left( h^{(1)}_{k_\mathrm{cts}+1,{n+1}}
                - h^{(1)}_{k_\mathrm{cts},{n+1}} \right)\,,
  \label{eq:freez_trans_discr_1}
\end{equation}
and the enthalpy gradient on the cold side of the CTS is
discretized by
\begin{equation}
  {\left( \frac{\partial h}{\partial z} \right)\!\!}^+_{n+1}
  \sim \frac{h^{(2)}_{k_\mathrm{cts}+1,{n+1}} 
             - h^{(2)}_{k_\mathrm{cts},{n+1}}}{\Delta z}\,.
  \label{eq:freez_trans_discr_2}
\end{equation}
Inserting Eqs.~(\ref{eq:freez_trans_discr_1}) and
(\ref{eq:freez_trans_discr_2}) in Eq.~(\ref{eq:freezcond})$_2$ yields
the updated volume flux $(a_\mathrm{m}^\perp)_{n+1}$. According to
Eqs.~(\ref{eq:velcts}) and (\ref{eq:cts_position_update}), this allows
to update subsequently the kinematic velocity of the CTS, $(w_z)_{n+1}$,
and the CTS position, $M_{n+1}$. The latter actually constitutes a
sub-grid tracking of the CTS that goes beyond the grid-limited precision
provided by $k_\mathrm{cts}$ (uppermost temperate grid point).

The final, updated enthalpy profile $h_{k,n+1}$ is assembled from
$h^{(1)}_{k,n+1}$ and $h^{(2)}_{k,n+1}$ according to
Eq.~(\ref{eq:enthalpy_profile_assembly}).

\section{Set-up of the numerical experiments}
\label{sec:setup}

For all numerical experiments presented in this work, the thickness of
the polythermal slab is $H=200\,\mathrm{m}$ and the inclination angle
$\gamma=4^\circ$ (Fig.~\ref{fig:twolayerslab}). For melting conditions
at the CTS, we set $v_z=-0.2\,\mathrm{m\,a^{-1}}$, and for freezing
conditions $v_z=+0.2\,\mathrm{m\,a^{-1}}$. All experiments are designed
such that the condition $|w_z|<|v_z|$ (see Eq.~(\ref{eq:volumeflux})
and the following text) is fulfilled at all times.

Steady states for enthalpy profiles, or, equivalently, for temperature
and water-content profiles, are computed for melting conditions with
surface temperatures of $-1^\circ$C and $-3^\circ$C, and for freezing
conditions with $-6^\circ$C and $-10^\circ$C.

Transient experiments are carried out for melting conditions with step
changes of the surface temperature from $-4^\circ$C to $-2^\circ$C and
vice versa, and for freezing conditions with step changes from
$-10^\circ$C to $-6^\circ$C and vice versa. The step changes are
employed at the initial time $t_\mathrm{init}=0$, and the initial conditions
for these four scenarios are steady states for the respective initial
temperatures.

Furthermore, we perform experiments with sinusoidal variations of the
surface temperature with a mean value of $-2^\circ$C and an amplitude of
1\,K for a melting CTS, and a mean value of $-8^\circ$C and an amplitude
of 2\,K for a freezing CTS. The initial conditions for these
scenarios are steady states for the mean surface temperatures
($-2^\circ$C and $-8^\circ$C, respectively), and two different periods of
100 and 500 years are employed for both cases.

All discussed experiments with the two-layer front-tracking scheme are
run with a resolution of 100 grid points in each layer and a time step
of 0.01 years. The standard resolution and time step for experiments
with the three different one-layer schemes are 1\,m (that is, 200 grid
points) and 0.01 years, respectively. However, we also use the
combinations 2\,m (100 grid points) / 0.01 years and 0.5\,m (400 grid
points) / 0.002 years; this is indicated in the captions of the
corresponding figures. All of these combinations are stable for both the
explicit and implicit versions of the several numerical schemes. The
implicit schemes allow for larger time steps of up to 100 years;
however, then the accuracy of transient solutions is affected. As long
as small time steps (within the range of stability of the explicit
schemes) are used, differences of results between the explicit and
implicit schemes are very small and virtually indistinguishable in
normal plots. Therefore, we only show results computed with the explicit
schemes.

\section{Results}
\label{sec:results}

\subsection{Two-layer front-tracking scheme}
\label{ssec:results_two_layer}

\begin{figure}[htb]
\begin{center}
\includegraphics[width=100mm]{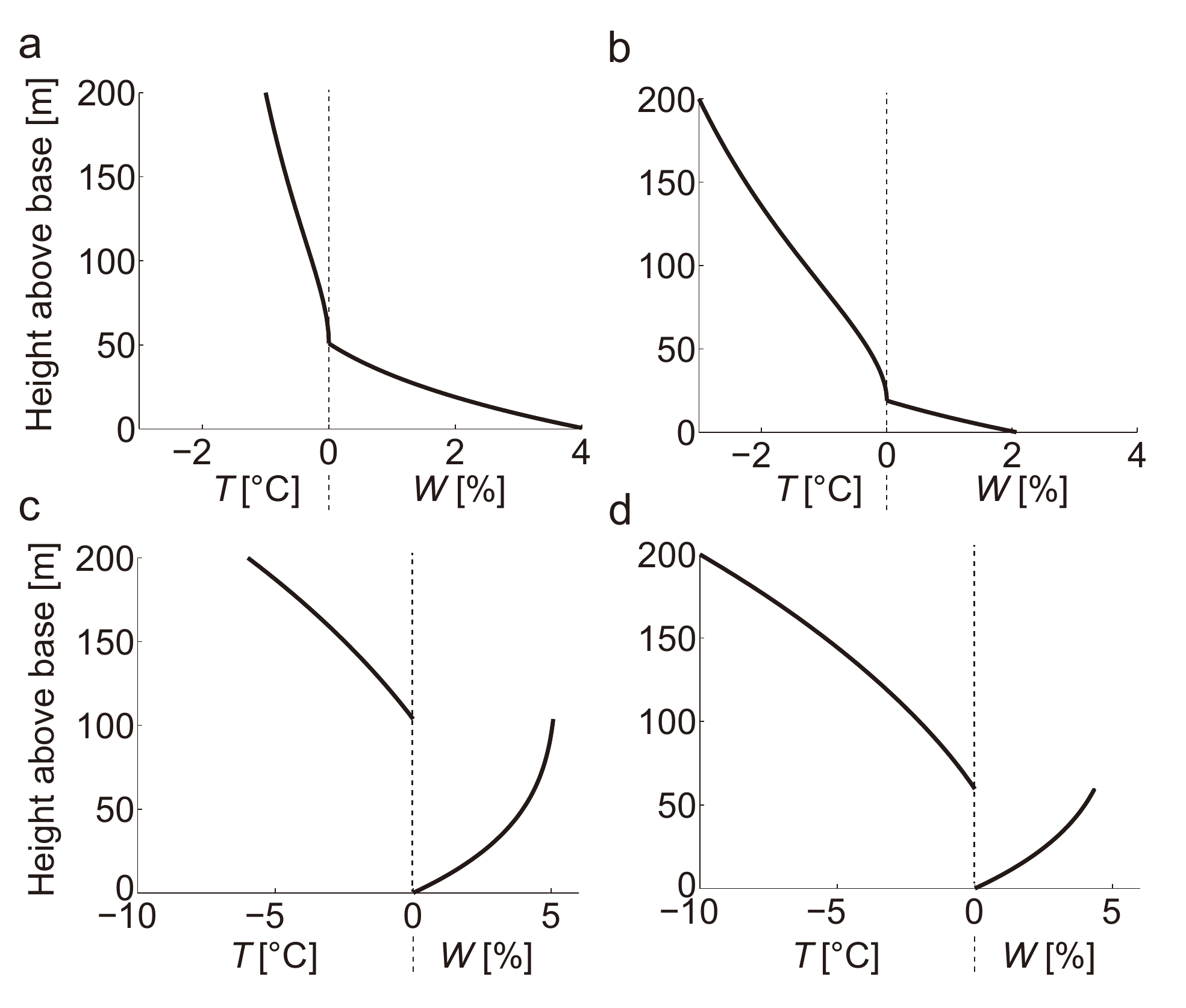}
\end{center}
\caption{Steady-state profiles of temperature $T$
(in the cold layer, $\mbox{values}\le{}0$) and water content $W$
(in the temperate layer, $\mbox{values}\ge{}0$)
of the parallel-sided slab
with a melting CTS ($v_z=-0.2\,\mathrm{m\,a^{-1}}$),
computed with the two-layer front-tracking scheme.
(a) surface temperature $T_\mathrm{s}=-1^\circ$C,
(b) $T_\mathrm{s}=-3^\circ$C.
Same for a freezing CTS ($v_z=+0.2\,\mathrm{m\,a^{-1}}$)
with (c) surface temperature $T_\mathrm{s}=-6^\circ$C,
(d) $T_\mathrm{s}=-10^\circ$C.}
\label{fig:twolayermelt}
\end{figure}

The steady-state results of the two-layer front-tracking tracking scheme
can be verified with the exact, analytical solutions \citep{Greve1997,
GreveBlatter2009}. Figure~\ref{fig:twolayermelt} shows the computed
steady-state solutions (enthalpy profiles converted back to temperature
and water-content profiles) for both a melting and freezing CTS and
prescribed constant surface enthalpies $h_\mathrm{s}=c\,T_\mathrm{s}$
corresponding to the surface temperatures listed in
Section~\ref{sec:setup}. These steady states agree to high accuracy with
the exact solutions (not shown explicitly). The positions of the CTS
coincide within about 0.3\,m, which is better than the grid resolution.
In particular, according to its design, the scheme handles the
discontinuities for freezing conditions at the CTS well.

We also calculated the various transient scenarios (step changes
in the surface boundary conditions, sinusoidally varying surface
conditions; see Section~\ref{sec:setup}) with the two-layer
front-tracking tracking scheme. The results (CTS positions and
maximum water contents as functions of time for the step changes,
CTS positions only for the sinusoidal forcings) are shown below in
Figs.~\ref{fig:meltstep}-\ref{fig:freeze-periodic} and will be
used as references to test the performance of the one-layer
schemes.

All runs were performed with both the explicit and the implicit
version of the two-layer front-tracking tracking scheme. The
steady-state solutions are almost independent of the time step
(within stability limits).
For the transient scenarios, time steps longer than the standard
time step (Section~\ref{sec:setup}) act like a low-pass filter
(again, within stability limits).
They have little influence on the results as long as the time step
is smaller than the rate of change of the conditions. Otherwise, 
delayed responses and dampenings occur, which, for the sinusoidal
forcings, results in reduced amplitudes and phase shifts of the
oscillating solutions.

\subsection{Conventional one-layer scheme}
\label{ssec:results_one_layer_1}

\begin{figure}[htb]
\begin{center}
\includegraphics[width=75mm]{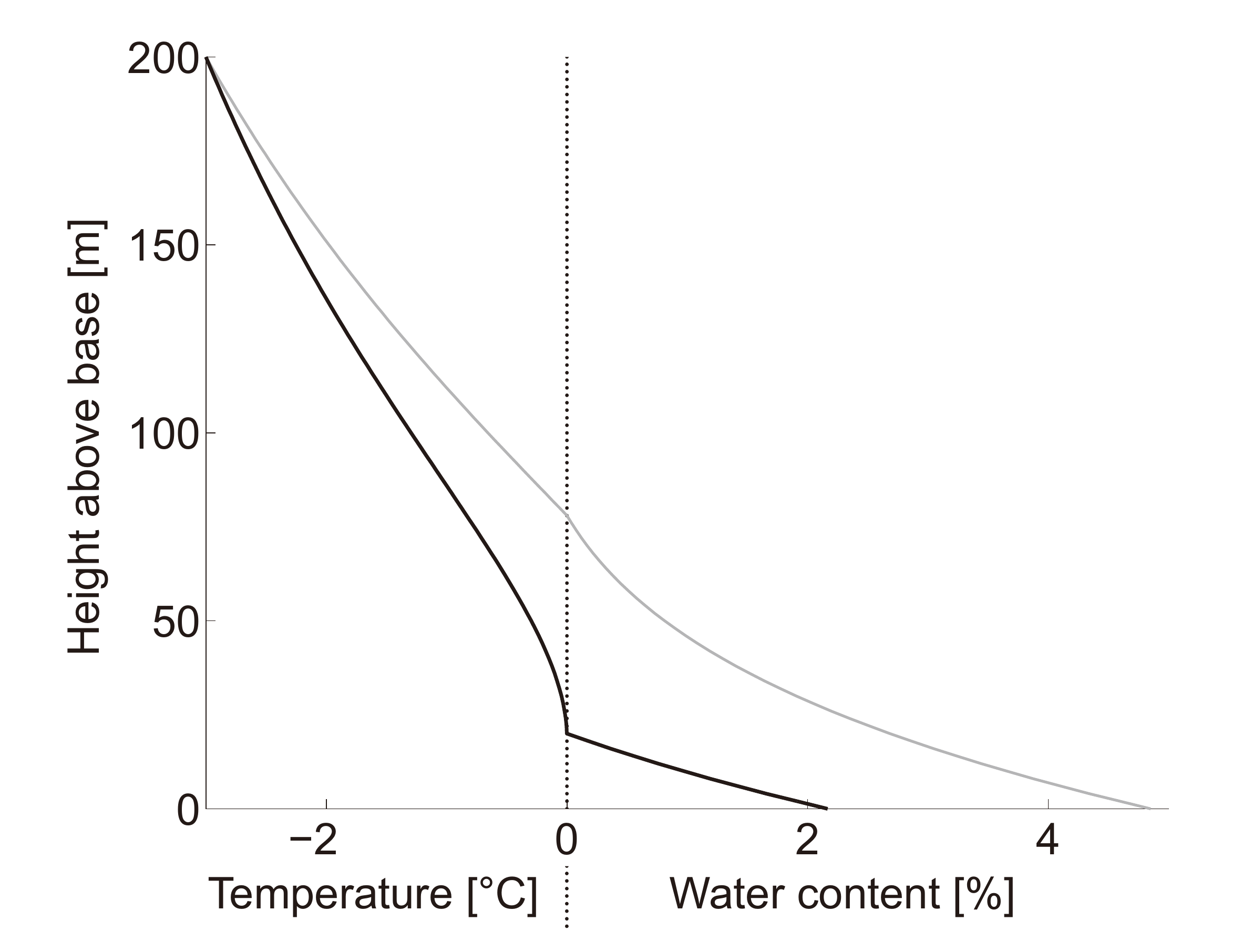}
\end{center}
\caption{Black, lower line: Steady-state profiles of temperature 
(in the cold layer, $\mbox{values}\le{}0$) and water content
(in the temperate layer, $\mbox{values}\ge{}0$)
corresponding to the steady-state solution shown in
Fig.~\ref{fig:twolayermelt}b for a melting CTS
($v_z=-0.2\,\mathrm{m\,a^{-1}}$), computed with the
conventional one-layer scheme.
Grey, upper line: Same, but the jump of the conductivity at the CTS
was disregarded in the discretization of the diffusion term
in Eq.~(\ref{eq:field-enthalpy-time}).}
\label{fig:gridstepenthalpy}
\end{figure}

Figure \ref{fig:gridstepenthalpy} (black line) shows the steady-state
solution (enthalpy profile converted back to temperature and
water-content profiles) for a melting CTS and a surface enthalpy
corresponding to $T_\mathrm{s}=-3^\circ$C computed with the conventional
one-layer scheme, which corresponds to the enthalpy scheme by
\citet{AschwandenBueler2012}. The solution is almost identical to the
exact solution and the one computed with the two-layer
front-tracking scheme (Fig.~\ref{fig:twolayermelt}b).

However, this
result is only obtained if the jump of the conductivity at the CTS is
properly accounted for in the discretization of the diffusion term in
the enthalpy equation (Eq.~(\ref{eq:discr_across_cts})). 
Otherwise (discontinuity of the conductivity at the CTS disregarded,
discretization like in Eq.~(\ref{eq:discr_const_kappa})), a greatly
flawed solution results (grey line in Fig.~\ref{fig:gridstepenthalpy}).
This solution has significantly larger enthalpies along the entire
profile except for the surface, the CTS position is 
$\sim{}4$ times higher above the base than in the correct solution,
and it does not meet the required transition condition
(\ref{eq:meltcond})$_2$ at the CTS.

\subsection{One-layer melting CTS scheme}
\label{ssec:results_one_layer_2}

\begin{figure}
\begin{center}
\includegraphics[width=80mm]{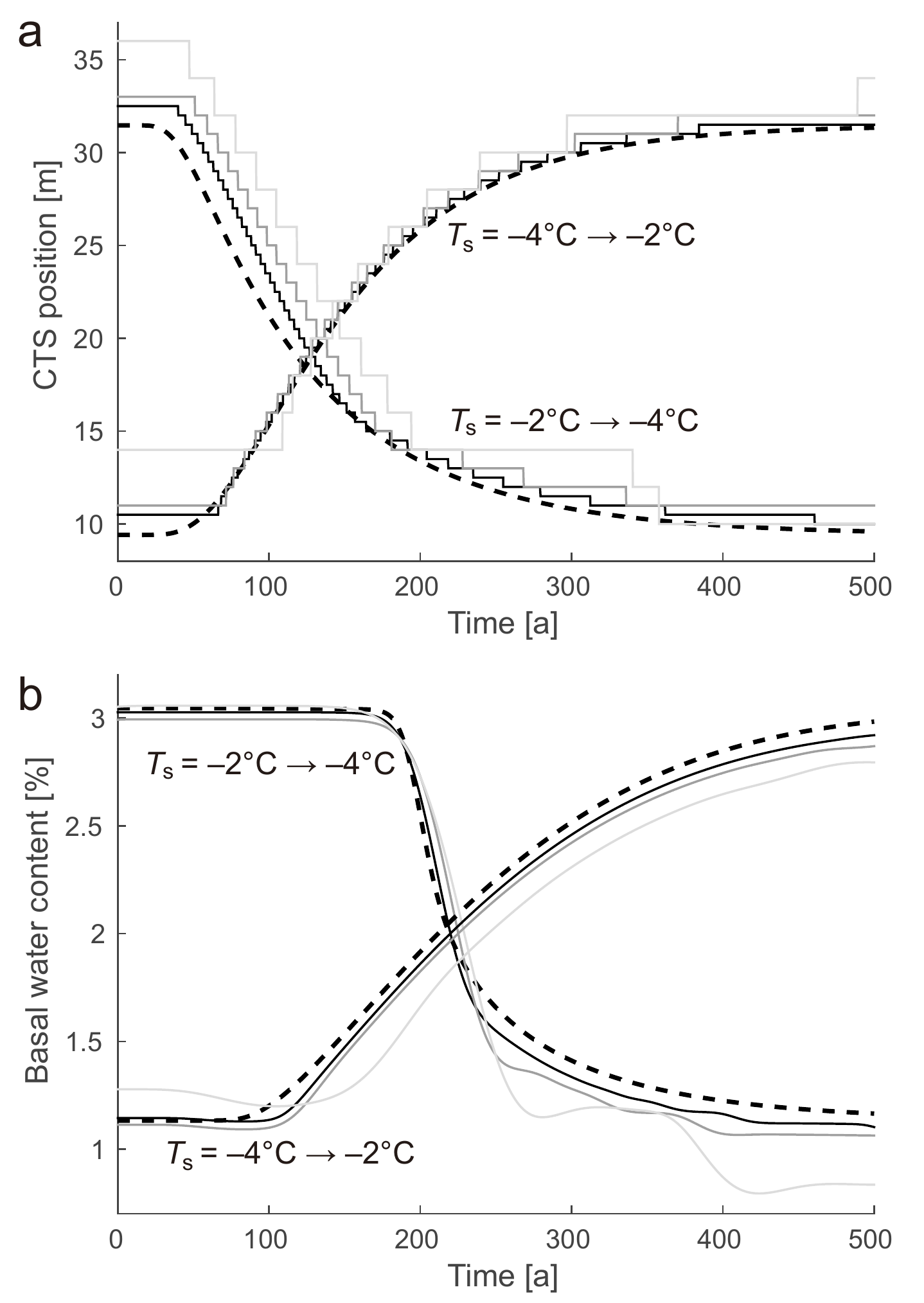}
\end{center}
\caption{Comparison between the one-layer melting CTS scheme (solid
lines) and the two-layer front-tracking scheme (dashed lines)
for melting conditions ($v_z=-0.2\,\mathrm{m\,a^{-1}}$).
(a) Position of the CTS and
(b) basal water content for a step change of the surface temperature
from $T_\mathrm{s}=-4^\circ$C to $-2^\circ$C (rising curves)
and vice versa (falling curves) at time $t=0$.
The three different solid lines correspond to grid resolutions
of 0.5\,m (black), 1\,m (medium-grey) and 2\,m (light-grey).}
\label{fig:meltstep}
\end{figure}

Figure \ref{fig:meltstep} shows a comparison of the one-layer melting
CTS scheme with the two-layer front-tracking scheme that is considered
to provide reference solutions. It shows the position of the CTS
and the basal water content for the step-change scenarios from
$T_\mathrm{s}=-4^\circ$C to $-2^\circ$C and vice versa
(Section~\ref{sec:setup}). For the one-layer scheme, the three
different resolutions of 0.5\,m, 1\,m (standard) and 2\,m have
been employed.

The transitions between the two states show some asymmetric behaviour,
depending on whether the CTS moves in the direction of the cold or
the temperate layer. The evolution of the CTS position is smooth
for the two-layer scheme, whereas it occurs in steps for the
one-layer scheme. This is a consequence of the CTS tracking on
the discrete grid in the one-layer scheme
(Section~\ref{ssec:numerics_one_layer_2}), which only allows an
accuracy of one grid spacing. The results obtained with the one-layer
scheme (CTS position, basal water content) show good convergence with
increasing resolution to the reference solution obtained with the
two-layer scheme. 

\begin{figure}
\begin{center}
\includegraphics[width=80mm]{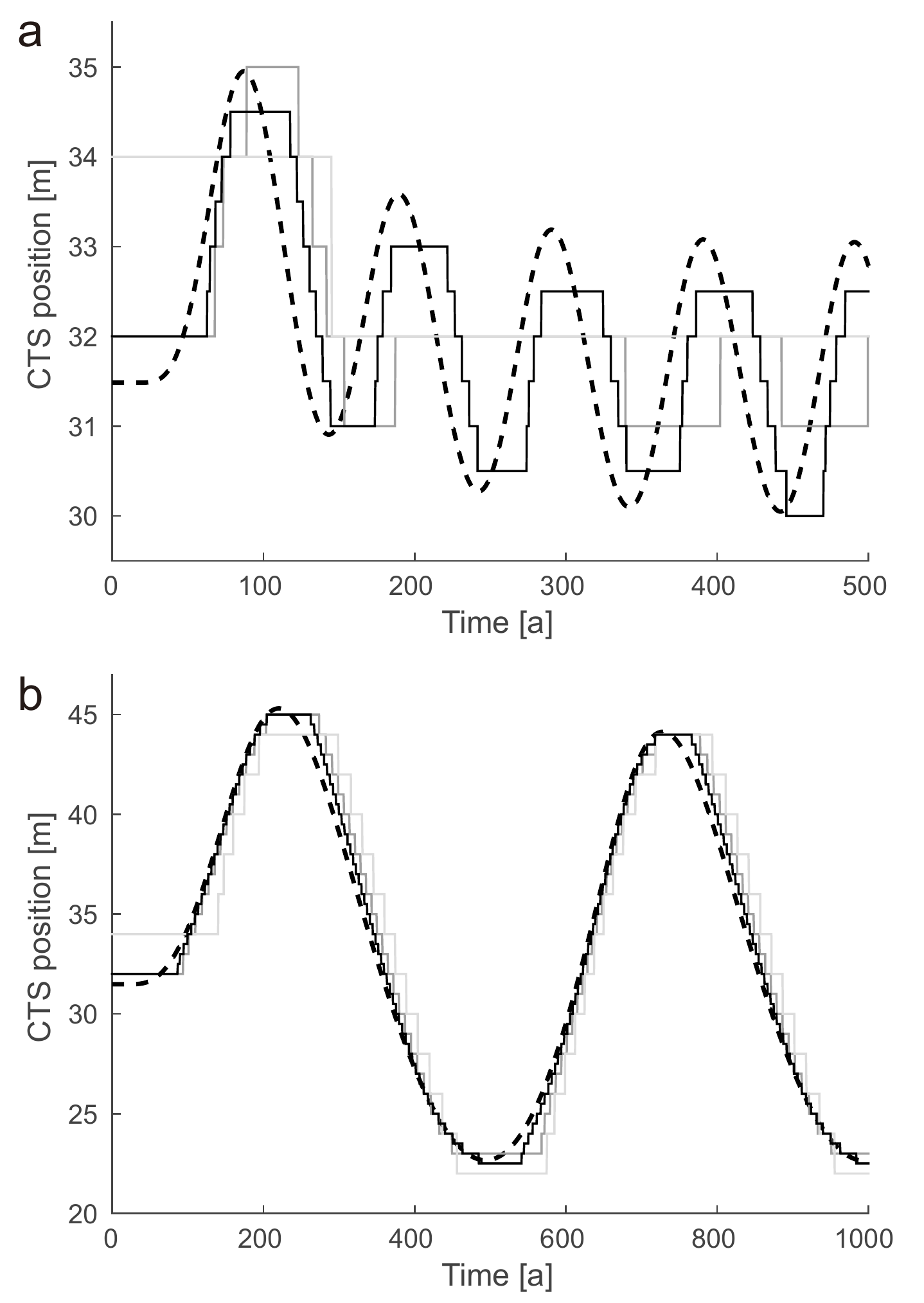}
\end{center}
\caption{Comparison between the one-layer melting CTS scheme
(solid lines) and the two-layer front-tracking scheme (dashed lines)
for melting conditions ($v_z=-0.2\,\mathrm{m\,a^{-1}}$).
(a) Position of the CTS for a sinusoidal oscillation of the
surface temperature centred at $T_\mathrm{s}=-2^\circ$C with an
amplitude of 1\,K and a period of 100 years.
(b) same as (a), but with a period of 500 years.
The three different solid lines correspond to grid resolutions
of 0.5\,m (black), 1\,m (medium-grey) and 2\,m (light-grey).}
\label{fig:sigmastep_periodic}
\end{figure}

The computed CTS positions for the two scenarios with sinusoidal
forcings (Section~\ref{sec:setup}) are shown in 
Fig.~\ref{fig:sigmastep_periodic}. Again, the results obtained
with the two-layer scheme are very smooth, while the one-layer
scheme produces step changes of the CTS position that reflect
the resolution. This becomes particularly evident for the short 
period (100 years), for which the amplitude of the variability
of the CTS position is less then 2\,m. It can only be reproduced
reasonably well by the highest resolution of 0.5\,m, while the
1-m resolution reproduces the variability only rudimentarily,
and the 2-m resolution yields only an average CTS position.
For the long period (500 years), the results are less susceptible
to the grid resolution, of comparable quality to those for
the step-change forcing shown in Fig.~\ref{fig:meltstep}a, and
the highest 0.5-m resolution matches closely the solution for
the two-layer front-tracking method.

\begin{figure}
\begin{center}
\includegraphics[width=80mm]{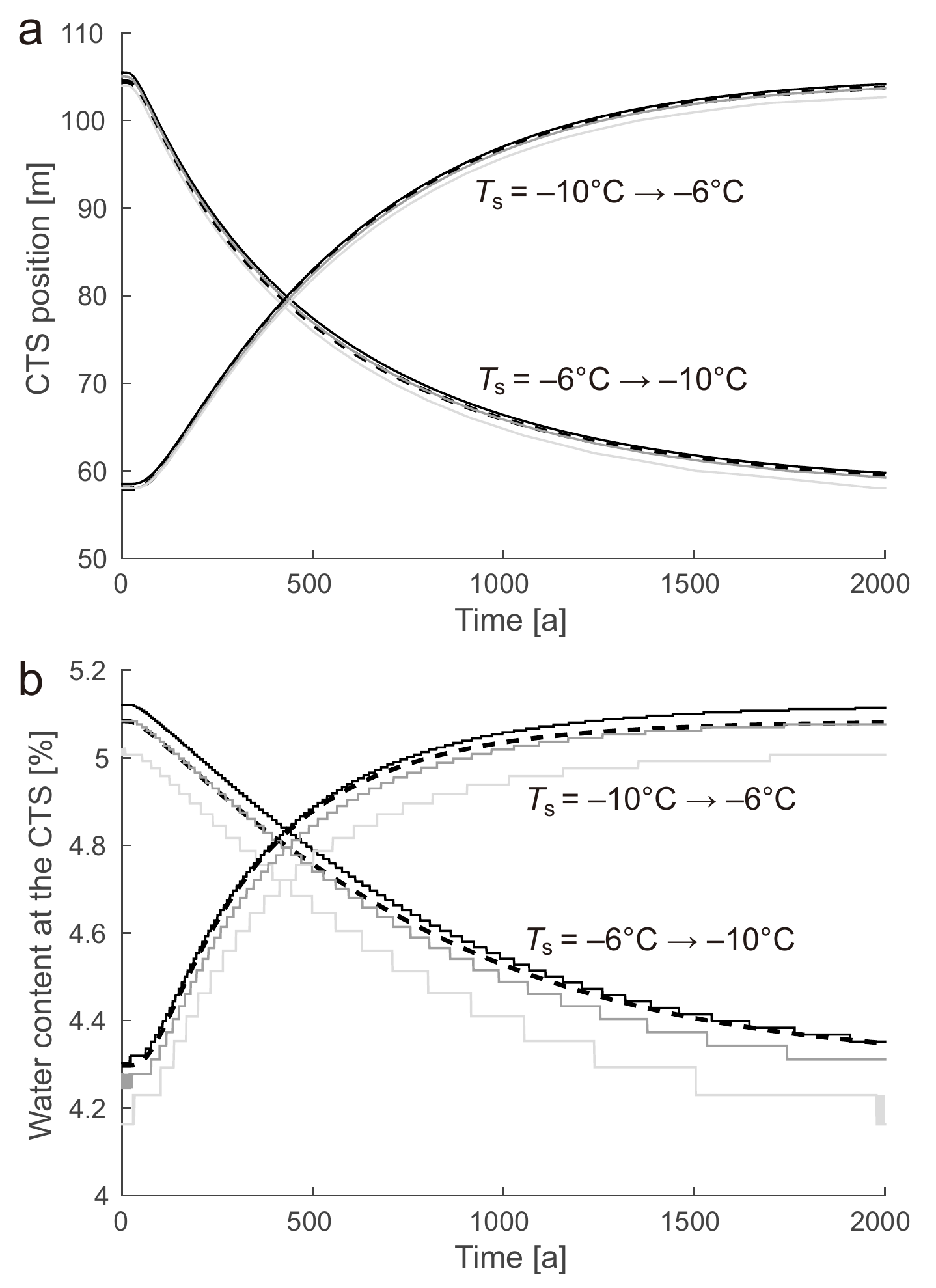}
\end{center}
\caption{Comparison between the one-layer freezing CTS scheme (solid
lines) and the two-layer front-tracking scheme (dashed lines)
for freezing conditions ($v_z=+0.2\,\mathrm{m\,a^{-1}}$).
(a) Position of the CTS and
(b) water content at the temperate side of the CTS
for a step change of the surface temperature
from $T_\mathrm{s}=-10^\circ$C to $-6^\circ$C (rising curves)
and vice versa (falling curves) at time $t=0$.
The three different solid lines correspond to grid resolutions
of 0.5\,m (black), 1\,m (medium-grey) and 2\,m (light-grey).}
\label{fig:freezestep}
\end{figure}

\begin{figure}
\begin{center}
\includegraphics[width=80mm]{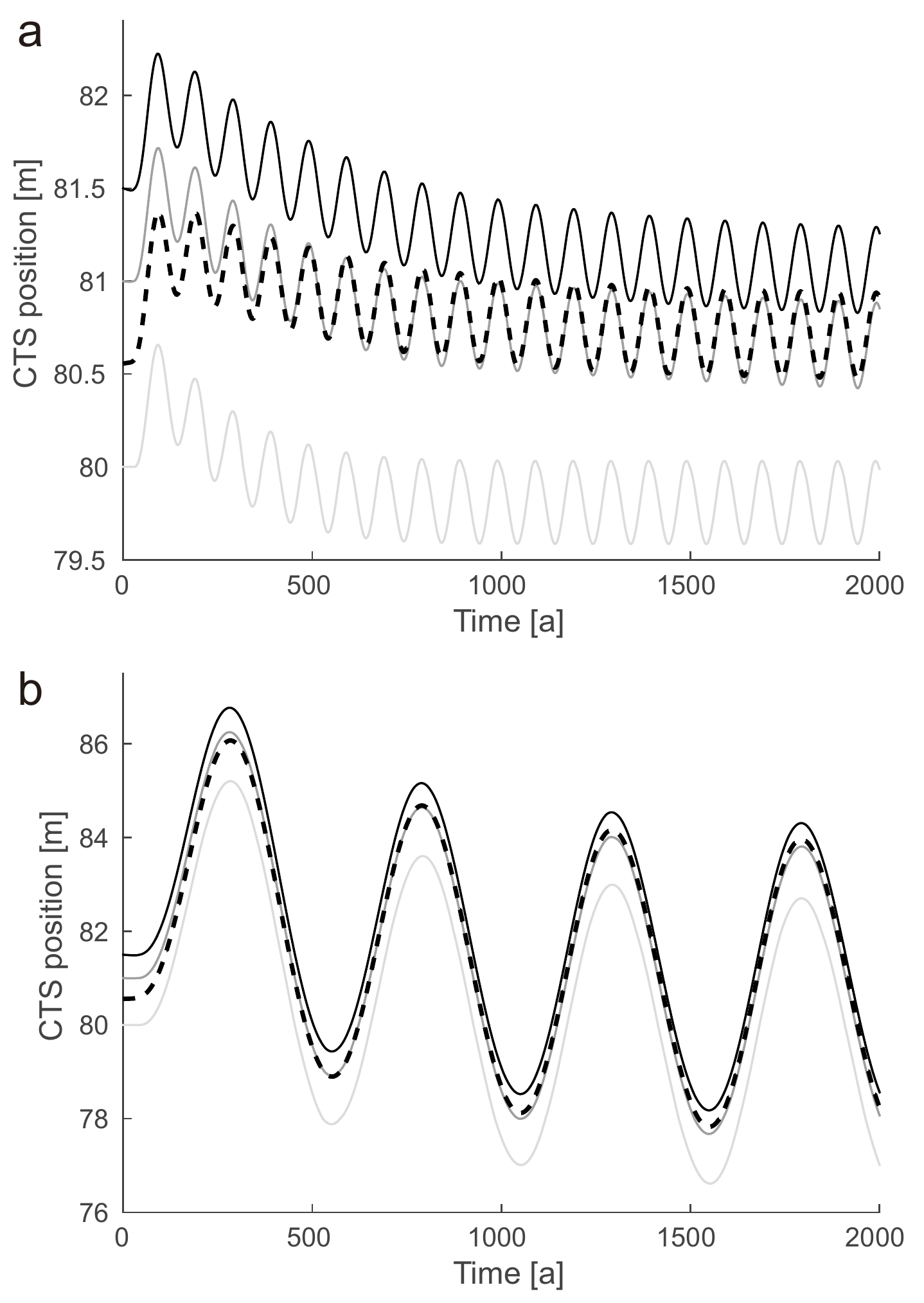}
\end{center}
\caption{Comparison between the one-layer freezing CTS scheme
(solid lines) and the two-layer front-tracking scheme (dashed lines)
for freezing conditions ($v_z=+0.2\,\mathrm{m\,a^{-1}}$).
(a) Position of the CTS for a sinusoidal oscillation of the
surface temperature centred at $T_\mathrm{s}=-8^\circ$C with an
amplitude of 2\,K and a period of 100 years.
(b) same as (a), but with a period of 500 years.
The three different solid lines correspond to grid resolutions
of 0.5\,m (black), 1\,m (medium-grey) and 2\,m (light-grey).}
\label{fig:freeze-periodic}
\end{figure}

\subsection{One-layer freezing CTS scheme}
\label{ssec:results_one_layer_3}

We now compare the performance of the one-layer freezing CTS scheme
with that of the two-layer front-tracking scheme.
Figure \ref{fig:freezestep} shows the evolution of the height of the
CTS above the bed and the water content at the temperate side of the
CTS for the step-change scenarios from
$T_\mathrm{s}=-10^\circ$C to $-6^\circ$C and vice versa
(Section~\ref{sec:setup}).
Like in Section~\ref{ssec:results_one_layer_2}, for the one-layer scheme,
the three different resolutions of 0.5\,m, 1\,m (standard) and 2\,m have
been employed.

Since the upward-moving ice reduces the response of the system to
surface perturbations at a given depth, the freezing CTS requires
a longer time for adjustment than the melting CTS
(compare with Fig.~\ref{fig:meltstep}a). A further, notable
difference is that the CTS evolution is smooth for both the
one-layer and two-layer schemes. This is a consequence of the
sub-grid tracking of the CTS employed in the one-layer freezing
CTS scheme (Section~\ref{ssec:numerics_one_layer_3}), whereas the
one-layer melting CTS scheme allows tracking the CTS only with
grid-limited precision
(Section~\ref{ssec:numerics_one_layer_2}).
By contrast, the computed water content
at the CTS shows some step changes for the one-layer scheme,
while it is also smooth for the two-layer scheme. For all three
resolutions of the one-layer scheme, the results agree well
with those of the two-layer scheme; the largest (but still
acceptable) discrepancy is found for the water content at the
CTS computed with the 2-m resolution.

Figure \ref{fig:freeze-periodic} shows the evolutions of the CTS
positions for the two scenarios with sinusoidal forcings
(Section~\ref{sec:setup}), computed with the one-layer freezing CTS
scheme and the two-layer scheme. The amplitude of the variations of the
CTS position is substantially smaller than that of the melting CTS even
though the amplitude of the surface perturbation is larger (2\,K vs.\
1\,K), which is again due to the upward direction of the ice motion that
delays and dampens changes of the surface conditions at depth. For both
periods (100 and 500 years) and all resolutions, the agreement to the
reference results (two-layer scheme) is within about the grid resolution
of the one-layer scheme. As already observed in
Fig.~\ref{fig:freezestep}a, the sub-grid tracking leads to a smooth
evolution of the CTS, and the simulated amplitudes agree very well for
all schemes and resolutions.

\section{Discussion and conclusion}
\label{sec:discussion}

The conventional one-layer scheme, which corresponds to the
implementation of the enthalpy method by \citet{AschwandenBueler2012},
does not explicitly take into account the Stefan-type energy- and
mass-flux matching conditions at the CTS that are crucial for
determining the position of the CTS. Nevertheless, we have demonstrated
that this scheme can determine the CTS position for the case of melting
conditions correctly. However, this depends critically on the
proper numerical handling of the discontinuity of the conductivity
across the CTS and is therefore prone to failure if the
implementation is not done with great care. For the case of freezing
conditions with the associated discontinuity of the enthalpy at the CTS,
the conventional one-layer scheme fails inevitably.

Two-layer front-tracking schemes, using a time-dependent
terrain-fol\-low\-ing coordinate transformation for the cold and temperate
layers separately \citep{Blatter1991,Greve1997,Pettersson2007}, can also
be used in conjunction with the enthalpy method. We have constructed
such a scheme, and have shown that it produces very good results for
both melting and freezing conditions. However, schemes using only one
grid for the entire polythermal slab are simpler to implement in
existing ice sheet models and therefore more desirable.

Therefore, we have proposed one-layer methods that modify the original
enthalpy scheme by \citet{AschwandenBueler2012} in order to treat
explicitly the transition conditions at the CTS for both cases of a
melting and freezing CTS. The proposed methods work well in our
one-dimensional model, provided that the time steps and grid resolutions
are sufficiently small. We expect them to work as well in shallow ice
sheet models, where the thermodynamics neglects horizontal diffusive
heat fluxes and thus treats vertical enthalpy or temperature profiles
essentially in a one-dimensional way. Horizontal advective heat fluxes
can be treated as additional source terms of the vertical profiles. In
fact, we have already implemented the one-layer melting CTS scheme in
the ice sheet model SICOPOLIS (e.g.,
\citeauthor{sato_greve_2012_annglac},
\citeyear{sato_greve_2012_annglac};
\citeauthor{greve_herzfeld_2013_annglac},
\citeyear{greve_herzfeld_2013_annglac}; URL http://www.sicopolis.net/),
which, despite the required adjustments for the additional physics
(pressure dependence of the melting point as well as
temperature-dependent heat conductivity and specific heat capacity
accounted for), could be done in a fairly straightforward way (paper in
preparation). With some additional effort due to the complicating
horizontal diffusive heat fluxes, implementations in non-shallow
(higher-order or full Stokes) ice sheet and glacier models should also
be feasible.

\section*{Acknowledgements}

We thank F.\ Saito, A.\ Aschwanden, E.\ Bueler and T.\ Kleiner
for helpful discussions. 
Comments by two anonymous reviewers and the scientific editor,
T.~Ka\-me\-da, helped to improve the manuscript.
H.B.\ was supported by an Invitation Fellowship for Research in Japan
(No.\ L13525) of the Japan Society for the Promotion of Science (JSPS).
R.G.\ was supported by a JSPS Grant-in-Aid for Scientific Research A
(No.\ 22244058).



\begin{thebibliography}{15}
\expandafter\ifx\csname natexlab\endcsname\relax\def\natexlab#1{#1}\fi
\expandafter\ifx\csname url\endcsname\relax
  \def\url#1{\texttt{#1}}\fi
\expandafter\ifx\csname urlprefix\endcsname\relax\def\urlprefix{URL }\fi

\bibitem[{Aschwanden and Blatter(2005)}]{Aschwanden2005}
Aschwanden, A., Blatter, H., 2005. {Meltwater production due to strain heating
  in Storglaci\"{a}ren, Sweden}. J. Geophys. Res. 110~(F4), F04024.

\bibitem[{Aschwanden et~al.(2012)Aschwanden, Bueler, Khroulev, and
  Blatter}]{AschwandenBueler2012}
Aschwanden, A., Bueler, E., Khroulev, C., Blatter, H., 2012. An enthalpy
  formulation for glaciers and ice sheets. J. Glaciol 58~(209), 441--457.

\bibitem[{Blatter and Hutter(1991)}]{Blatter1991}
Blatter, H., Hutter, K., 1991. {Polythermal conditions in Artic glaciers}. J.
  Glaciol. 37~(126), 261--269.

\bibitem[{Dash et~al.(2006)Dash, Rempel, and Wettlaufer}]{Dash&al2006}
Dash, J.~G., Rempel, A.~W., Wettlaufer, J.~S., 2006. The physics of premelted
  ice and its geophysical consequences. Rev. Mod. Phys. 78~(3), 695--741.

\bibitem[{Duval(1977)}]{Duval1977}
Duval, P., 1977. The role of water content on the creep of polycrystalline ice.
  In: Isotopes and Impurities in Snow and Ice -- Proceedings of the Grenoble
  Symposium, August-September 1975. IAHS Publication No.~118. International
  Association of Hydrological Sciences, Wallingford, UK, pp. 29--33.

\bibitem[{Fowler(1984)}]{Fowler1984}
Fowler, A.~C., 1984. On the transport of moisture in polythermal glaciers.
  Geophys. Astrophys. Fluid Dyn. 28~(2), 99--140.

\bibitem[{Fowler and Larson(1978)}]{Fowler1978}
Fowler, A.~C., Larson, D.~A., 1978. {Flow of polythermal glaciers: 1. Model and
  preliminary analysis}. Proc. R. Soc. Lond. A 363~(1713), 217--242.

\bibitem[{Greve(1997)}]{Greve1997}
Greve, R., 1997. {A continuum-mechanical formulation for shallow polythermal
  ice sheets}. Phil. Trans. R. Soc. Lond. A 355~(1726), 921--974.

\bibitem[{Greve and Blatter(2009)}]{GreveBlatter2009}
Greve, R., Blatter, H., 2009. Dynamics of Ice Sheets and Glaciers. Springer,
  Berlin, Germany, etc.

\bibitem[{Greve and Herzfeld(2013)}]{greve_herzfeld_2013_annglac}
Greve, R., Herzfeld, U.~C., 2013. Resolution of ice streams and outlet glaciers
  in large-scale simulations of the {Greenland} ice sheet. Ann. Glaciol.
  54~(63), 209--220.

\bibitem[{Hutter(1982)}]{Hutter1982}
Hutter, K., 1982. A mathematical model of polythermal glaciers and ice sheets.
  Geophys. Astrophys. Fluid Dyn. 21~(3-4), 201--224.

\bibitem[{Hutter(1993)}]{Hutter1993}
Hutter, K., 1993. {Thermo-mechanically coupled ice-sheet response -- cold,
  polythermal, temperate}. J. Glaciol. 39~(131), 65--86.

\bibitem[{Kleiner et~al.(2015)Kleiner, R{\"u}ckamp, Bondzio, and
  Humbert}]{Kleiner&al2015}
Kleiner, T., R{\"u}ckamp, M., Bondzio, J., Humbert, A., 2015. Enthalpy
  benchmark experiments for numerical ice sheet models. The Cryosphere 9~(1),
  217--228.

\bibitem[{Pettersson et~al.(2007)Pettersson, Jansson, Huwald, and
  Blatter}]{Pettersson2007}
Pettersson, R., Jansson, P., Huwald, H., Blatter, H., 2007. Spatial pattern and
  stability of the cold surface layer of {Storglaci\"aren, Sweden}. J. Glaciol.
  53~(180), 99--109.

\bibitem[{Sato and Greve(2012)}]{sato_greve_2012_annglac}
Sato, T., Greve, R., 2012. Sensitivity experiments for the {Antarctic} ice
  sheet with varied sub-ice-shelf melting rates. Ann. Glaciol. 53~(60),
  221--228.

\end{thebibliography}
\end{document}